\documentclass[journal=jpclcd,manuscript=letter]{achemso}
\usepackage{stix}
\usepackage[utf8]{inputenc}
\usepackage{amsmath}
\usepackage{bm}
\usepackage{hyperref}
\usepackage{graphicx}
\usepackage{comment}
\usepackage{enumitem}
\usepackage{verbatim}
\usepackage{multirow}
\usepackage{longtable}
\usepackage{array}

\usepackage{booktabs}
\usepackage[version=3]{mhchem} 
\newcommand{\beq}{\begin{eqnarray}}
\newcommand{\eeq}{\end{eqnarray}}
\usepackage[dvipsnames]{xcolor}
\usepackage[normalem]{ulem}
\usepackage{mathtools}
\SectionNumbersOn
\mciteErrorOnUnknownfalse

\DeclarePairedDelimiter\ket{\lvert}{\rangle}
\DeclarePairedDelimiterX\braOket[3]{\langle}{\rangle}{#1\,\delimsize\vert\,\mathopen{}#2\,\delimsize\vert\,\mathopen{}#3}
\DeclarePairedDelimiterX\braket[2]{\langle}{\rangle}{#1\,\delimsize\vert\,\mathopen{}#2}

\newcommand{\REV}[1]{\textcolor{black}{#1}}

\author{Linqing Peng}
\affiliation{Division of Chemistry and Chemical Engineering, California Institute of Technology, Pasadena CA 91125, USA}
\author{Shuanglong Liu}
\affiliation{Department of Physics, Northeastern University, Boston, Massachusetts 02115, USA}
\author{Xing Zhang}
\affiliation{Division of Chemistry and Chemical Engineering, California Institute of Technology, Pasadena CA 91125, USA}
\author{Xiao Chen}
\affiliation{Department of Physics, Northeastern University, Boston, Massachusetts 02115, USA}
\author{Chenghan Li}
\affiliation{Division of Chemistry and Chemical Engineering, California Institute of Technology, Pasadena CA 91125, USA}
\author{Hai-Ping Cheng}
\email{ha.cheng@northeastern.edu}
\affiliation{Department of Physics, Northeastern University, Boston, Massachusetts 02115, USA}
\author{Garnet Kin-Lic Chan}
\email{gkc1000@gmail.com}
\affiliation{Division of Chemistry and Chemical Engineering, California Institute of Technology, Pasadena CA 91125, USA}
\title{Accurate crystal field Hamiltonians of single-ion magnets at mean-field cost}

\begin{document}

\begin{tocentry}
  \centering
\includegraphics{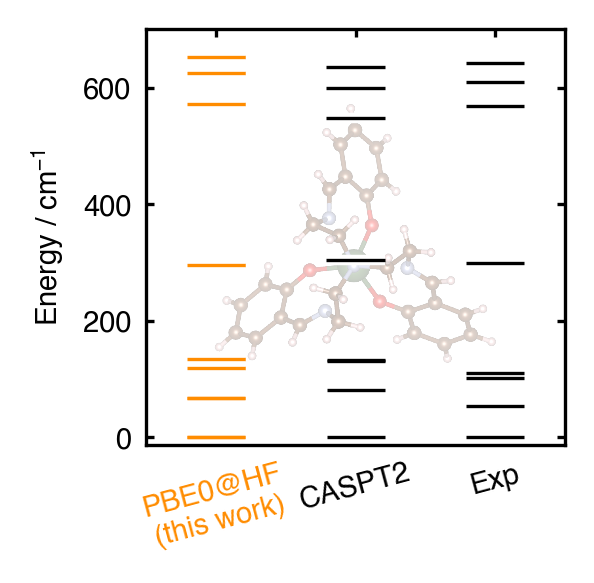}
\end{tocentry}

\begin{abstract}
The effective crystal field Hamiltonian provides the key description of the electronic properties of single-ion magnets, but obtaining its parameters from ab initio computation is challenging. 
We introduce a simple approach to derive the effective crystal field Hamiltonian through density functional calculations of randomly rotated mean-field states within the low-energy manifold.
In benchmarks on five lanthanide-based complexes, we find that we compute with mean-field cost an effective crystal field Hamiltonian that matches the state-of-the-art from much more expensive multi-configurational quantum chemistry methods. In addition, we are able to reproduce the experimental low-energy spectrum and magnetic properties with an accuracy exceeding prior attempts. Due to its low cost, our approach provides a crucial ingredient in the computational design of single-ion magnets with tailored physical properties and low-energy spectra.
\end{abstract}

\paragraph{Introduction}

Tuning the electronic properties of single-ion molecular complexes 
is a goal of synthetic chemistry for potential applications in areas of molecular magnetism~\cite{christou2000single, sessoli2017magnetic}, molecular  spintronics~\cite{bogani2008molecular, thiele2014electrically}, and in quantum information processing~\cite{gaita2019molecular,atzori2019second, bayliss2020optically}. 
In these systems, the effective crystal field Hamiltonian $\hat{H}_\text{CF}$ provides the theoretical rationalization of the low-energy properties, such as the excited states and electronic pathways for decoherence. For lanthanide complexes in particular, to which we will devote our attention in this work, $\hat{H}_\text{CF}$ describes the interplay between the strong spin-orbit coupling and associated zero-field splitting in the $f$-orbital shell and the crystal field, leading to a large number of effective parameters that need to be determined, especially in low-symmetry complexes. This complicates both experimental and theoretical procedures to accurately estimate $\hat{H}_\text{CF}$ in lanthanide complexes.

From a theory perspective, state-of-the-art procedures aim to determine multiple electronic eigenstates of the lanthanide complex which are subsequently fitted to the crystal field Hamiltonian~\cite{ungur2017ab,gomezcoca2015large, Atanasov2012modern}. However, the character of these eigenstates involving the partially filled 4$f$-shell often requires a sophisticated multi-configurational wavefunction description. Quantitative results can be obtained using the complete active space self-consistent field method (CASSCF)~\cite{roos1980complete,Malmqvist1989,Malmqvist2002} augmented by second-order perturbation theory such as complete active space perturbation theory (CASPT2)~\cite{andersson1990second,andersson1992second} or $n$-electron valence state perturbation theory (NEVPT2)~\cite{angeli2001introduction,angeli2001n}, or by multi-reference configuration interaction~\cite{buenker1974individualized, buenker1975energy,werner1988efficient}, but these are prohibitively expensive for many of the lanthanide-based single-ion complexes of experimental interest. \REV{Density functional theory (DFT) based methods~\cite{reviakine_calculation_2006,pederson_magnetic_1999,aquino_first-principle_2005,takeda_density_2005,neese_importance_2006,van_wullen_magnetic_2009,schmitt_zero-field_2011,kessler_broken_2013,neese_calculation_2007,kubica_zero-field_2013,khan_systematic_2015}, including both collinear and noncollinear spin formulations, have underestimated the splitting by a factor of 2 or more or even predicted the wrong sign, due to the limitations of the single-determinantal treatment of the eigenstates.}

In this work, we introduce a method to calculate the effective crystal field Hamiltonian for lanthanide-based single-ion complexes with an accuracy comparable to the state-of-the-art multireference theories, but at the cost of (several) mean-field Kohn-Sham density functional theory calculations. The basic insight is that although representing the eigenstates of $\hat{H}_\text{CF}$ might require a multireference treatment, obtaining the parameters of $\hat{H}_\text{CF}$ may not, and in particular by considering a sufficient number of single-reference states $\Psi_i$ (which need not be eigenstates) and their energies $E_i$, we can deduce the corresponding parameters of $\hat{H}_\text{CF}$. Similar techniques are used to derive effective Hamiltonians in other contexts, for example exchange parameters in spin Hamiltonians, which are often derived from the energies of broken symmetry spin states~\cite{schurkus}, and in the context of ligand field theory, the ligand field Hamiltonian has been derived in a similar procedure involving Slater determinants with both ground- and excited-occupancies~\cite{atanasov2005ground}. Here, we choose as our $\Psi_i$  single-configurational states that are optimized to (approximately) lie within the lowest spin-orbit coupled $|JM\rangle$ manifold using a variant of constrained density functional theory (cDFT)~\cite{kaduk2012constrained,behler2005dissociation,dederichs1984ground,behler2007nonadiabatic,wu2005direct} and we deduce the effective couplings of the crystal field Hamiltonian from their energies. \REV{A primary distinction from the work in Ref.~\cite{atanasov2005ground} is that the states whose energies we sample lie at the same energy scale as the eigenstates of the crystal field Hamiltonian, rather than spanning the much larger electronic energy scale of the different $f$-shell configurations.} We therefore refer to this as deriving the constrained DFT ab initio crystal field Hamiltonian. 

Our method captures both the dynamical and static effects of electron correlation on the crystal field Hamiltonian and, as we shall show below, achieves accuracy on par with state-of-the-art CASPT2- or NEVPT2- methods in small systems. However, as it has the same low computational scaling as density functional theory, it enables the derivation of the crystal field Hamiltonian in large molecules. 
As an example of the latter, we demonstrate our method on the holmium double--decker complex \REV{(113 atoms). Together with the recent application of our method to a new experimental lanthanide clock qubit~\cite{gakiya202554}, we see that we can achieve} an accuracy surpassing that of CASSCF and NEVPT2  in predicting both the energy spectrum and the magnetic properties. 

\paragraph{Constrained DFT ab initio crystal field Hamiltonian} 

\begin{figure}[htb!]
    \centering
    \includegraphics[width=0.5\columnwidth]{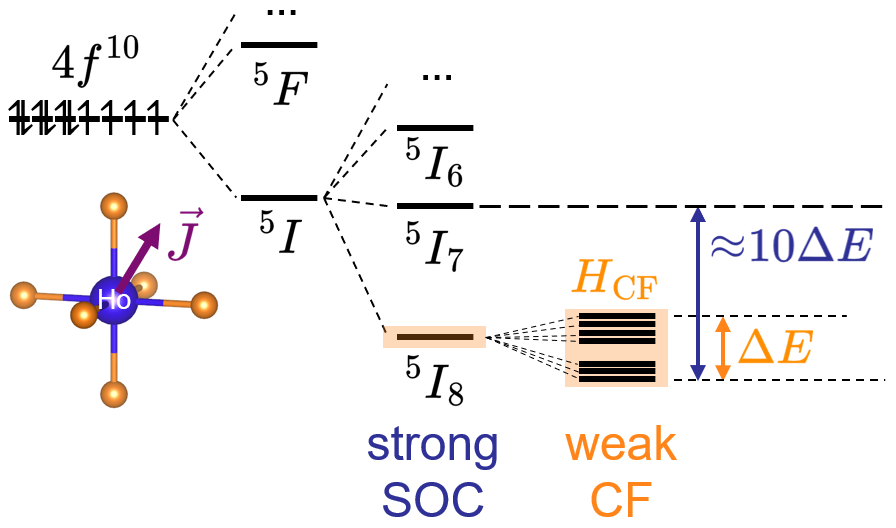}
    \caption{An illustration of the origin of the zero-field energy splitting (orange) in a Ho-based single-ion complex. Energy levels are not drawn to scale. }
    \label{figure:splitting}
\end{figure}

We will be interested in single-ion complexes where there is a strong spin-orbit coupling. At the two-component level of treating relativistic effects, this means that the electronic Hamiltonian  $\hat{H}$ contains a spin-orbit operator and the eigenstates are no longer eigenfunctions of spin. In addition to the spin-orbit coupling, which splits the non-relativistic atomic degeneracies, there is also the crystal field of the ligands. 
Fig.~\ref{figure:splitting} illustrates the structure of the low-energy manifold and the range of splittings for a typical  lanthanide complex.

The effective crystal field Hamiltonian is an operator in the space of the lowest manifold of states (the right of Fig.~\ref{figure:splitting}\REV{, the orange-shaded region})
\begin{align}
    \hat{H}_\mathrm{CF} = \sum_{JMJ'M'} h_{JM, J'M'} |JM\rangle \langle J'M'|
\end{align}
\REV{where $J$ is the total angular momentum quantum number of the lanthanide ion.} Assuming that these derive from a multiplet manifold from a single $J$ (which is the case when the SOC is  stronger than the crystal field splitting), the above effective Hamiltonian can also be written in terms of operators within this $|JM\rangle$ manifold, known as extended Stevens operator equivalents~\cite{stevens1952matrix}
\beq
\hat{H}_\mathrm{CF}(J) = \sum_{k=2,4,6}\sum_{q=-k}^{k}B_k^q \hat{O}_k^q(J)
\REV{=\sum_{k=2,4,6}\sum_{q=-k}^{k}B_k^q [O_k^q(J)]_{MM'} |JM\rangle \langle JM'|}
\eeq
where we use the definitions of $\hat{O}_k^q(J)$ in terms of $|JM\rangle \langle JM'|$ following the convention in Ref.~\cite{rudowicz1985transformation,rudowicz2004generalization,ryabov1999generation,onizhuk2021pycce}, \REV{and, depending on symmetry, the up to 27}  $B^q_k$ coefficients are the molecular-specific parameters to be determined.

Given a set of electronic low-energy states $\{ \Psi_i\}$ within the manifold on which $\hat{H}_\mathrm{CF}$ acts, we can determine the coefficients of $\hat{H}_\mathrm{CF}$ from 
\begin{align}
    E_i &= \mathrm{const} + \braOket{\Psi_i}{\hat{H}_\mathrm{CF}(J)}{\Psi_i}, \ i=1 \ldots D' \notag \\
    \Psi_i &= \sum_{JM} C_{JM, i} \Psi_i^{JM} \notag\\
    E_i &= \mathrm{const} + \sum_{JMJM'} C_{JM,i}^*C_{JM',i} \REV{B_k^q [O_k^q(J)]_{MM'}}
\label{eq:stevens}
\end{align}
where $\mathrm{const}$ is an energy zero, \REV{$C_{JM, i}$ is obtained by projecting the $i^{\mathrm{th}}$ low-energy state $\Psi_i$ into the eigenbases $\Psi_i^{JM}$ of $\hat{J}$ and $\hat{J}_z$ operators,
$[O_k^q(J)]_{MM'}$ are matrix elements of the extended Stevens operators in the eigenbases of  $\hat{J}$ and $\hat{J}_z$ operators as defined above,}  and the above constitutes a set of linear equations for the coefficients of $\hat{H}_\mathrm{CF}$\REV{, i.e. $B_k^q$} (if $D'=D$ where $D$ is the number of symmetry-allowed nonzero $B_k^q$ to be determined, generically we expect a unique solution; for $D'>D$, the linear equations can be solved in the least squares sense). The key to an efficient procedure is to use states $\{ \Psi_i \}$ for which the energy $E_i$ can be easily computed. Note that $\Psi_i$ need not be an eigenstate of the electronic Hamiltonian.
Instead, we can choose $\{ \Psi_i \}$ to be low-energy single-configurational states in the manifold of interest, whose energies can be accurately approximated by affordable electronic structure methods, such as density functional approximations.

For example, we can consider the case where $\hat{H}_\mathrm{CF}$ acts on a single $\{ |JM\rangle \}$ manifold where $J$ is the maximal $J$ associated with the given occupancy of the $f$-shell. 
The azimuthal projection angle is free to be chosen, and we denote the angular momentum state with projection axis $\vec{n}$, $|JM_{\vec{n}}\rangle$; then states of the form $\{ |J(M_{\vec{n}}=J)\rangle\}$, \REV{in which $\vec{J}$ is maximally parallel to $\vec{n}$,} are formed by a single $f$-shell Slater determinant composed of the set of generalized spin (spinor) orbitals with maximal projected angular momentum, i.e., $|j(m_{\vec{n}}=j)\rangle$, $|j(m_{\vec{n}}=j-1)\rangle$, etc., where the lowercase $j$, $m$ denote the single-orbital angular momenta. There is an infinite number of such single-determinant states, obtained by rotating the projection axis $\vec{n}$. {\REV{By uniformly sampling such states with a Haar random (uniformly distributed) $\vec{n}$, we generate a set of non-collinear solutions with different $\vec{J}$ orientations. 
This effectively samples the $|JM\rangle$ projections with equal probability.} 
We can then} use the energies of these states in Eq.~\ref{eq:stevens} \REV{to recover the $\hat{H}_\mathrm{CF}$ that describes the sampled subspace}.

In the general case, it may be necessary to partially expand the electronic manifold outside of the specific manifold of interest where $\hat{H}_\mathrm{CF}$ is defined, in order to construct low energy states of single-determinant character. For example, if $\hat{H}_\mathrm{CF}$ acts on a $\{ |JM\rangle \}$ manifold where $J$ is not the maximal $J$, then we may not find \REV{low-energy} 
single-determinant states within the $\{ |JM\rangle \}$ manifold (e.g., consider the case \REV{of Er$^{3+}$} where \REV{ground-state} $J=0$). However, by expanding the manifold to multiple $J$'s, $\{ |JM\rangle, |J'M'\rangle \}$, we can construct low-energy single determinant states which are not eigenstates of \REV{$\hat{J}^2$} (i.e. broken symmetry states) but with well defined $M_{\vec{n}}$ (equal to some specified $J$) \REV{(in the example of Er$^{3+}$, the low-energy determinant corresponding to $|L=3, M_L=3, S=3, M_S=-3\rangle$ consists of $|J=0, M_J=0\rangle$, $|J=1, M_J=0\rangle$, $|J=2, M_J=0\rangle$, etc., but has well defined $M_J=0$)}. These may be used to fit an effective $\hat{H}_\mathrm{CF}$ over the expanded manifold and the effective Hamiltonian can then be restricted to the single manifold of interest after the fact. 
\REV{If one wishes for additional control over the composition of the Slater determinant, we can also add penalizing terms that control the magnitude of $\vec{J}$ in addition to its orientation.}
This more general procedure is analogous to the use of broken-symmetry density functional and wavefunction calculations to determine effective Heisenberg parameters~\cite{noodleman,yamaguchi,schurkus}.

Formally, the Kohn-Sham density functional energies of the single-determinant states with $M_{\vec{n}} =J$ are defined through
\begin{align}
{E[\rho] = \min_{\Psi({M_{\vec{n}}=J})\to \rho} \left[\langle \Psi | \hat{h}_1 + \hat{V}_{\mathrm{eff}} | \Psi\rangle \right]}
\end{align}
where $\Psi$ is a Slater determinant with the azimuthal angular momentum constraint, 
$\hat{h}_1$ is the two-component one-electron part of the Hamiltonian in the given spin-orbit coupling treatment, and $\hat{V}_{\mathrm{eff}}$ includes the Coulomb, exchange, and correlation in  a general Kohn-Sham expressoin.
In practice, the constraint can be easily implemented by applying a penalty~\cite{ma2015constrained}, i.e., we minimize 
\begin{align}
     E[\rho, \vec{n}, \lambda] = \min_{\Psi \to \rho} \left[\langle \Psi | \hat{h}_1 + \hat{V}_{\mathrm{eff}} | \Psi\rangle  - \lambda(|\vec{J}| - \vec{n}\cdot \vec{J})\right]
\end{align}
\REV{where the total angular momentum vector $\vec{J}=(\langle \hat{J}_x\rangle, \langle \hat{J}_y\rangle, \langle \hat{J}_z\rangle)$ is a vector of expectation values of angular momentum operators $\hat{J}_x$, $\hat{J}_y$, and $\hat{J}_z$ evaluated using the lanthanide $f$-block of the generalized 1-particle reduced density matrix associated with the Kohn-Sham determinant.} Given such a formal definition, we therefore obtain the energy by minimizing the standard Kohn-Sham density functional energy with the additional penalty, and we choose $\lambda$ sufficiently large such that the constraint is well satisfied.

Because of the presence of different $J$ multiplets in an actual electronic structure calculation, the result of the above minimization is not a perfect $|J(M_{\vec{n}}=J)\rangle$ state, but contains a small admixture of other $J$ states. 
To expand the determinant into its $|JM\rangle$ components, we apply the group theoretic projection~\cite{percus1962exact} $C_{JM',i}\ket{\Psi_{JM,i}}=P_{M,M'}^J \ket{\Psi_i}$ where $C_{JM',i}=\braket{\Psi_{JM',i}}{\Psi_i}$, and the operator $P_{M,M'}^J$ is defined as 
\beq
P_{M,M'}^J=\frac{2J+1}{8\pi^2} \int d\Omega D_{M,M'}^{J*}(\Omega) \hat{R}_L(\Omega) \hat{R}_S(\Omega)
\eeq
where $\Omega=(\alpha, \beta, \gamma)$ are the Euler angles in z-y-z convention, $\hat{R}_L$ is the orbital rotation operator $\hat{R}_L=e^{-i\alpha \hat{L}_z} e^{-i\beta \hat{L}_y} e^{-i\gamma \hat{L}_z}$, $\hat{R}_S$ is the similarly defined spin rotation operator, and $D_{M,M'}^J(\Omega)=\braOket{JM}{\hat{R}(\Omega)}{JM'}$ is the Wigner $D$-matrix. The integration in the above can be efficiently computed by quadrature. Once the $|JM\rangle$ coefficients are obtained, we can solve Eq.~\ref{eq:stevens} for $h_{JM, J'M'}$ by least squares. We refer to the effective crystal field Hamiltonian derived this way as the constrained DFT ab initio crystal field Hamiltonian.

\paragraph{Computational implementation and details}

We have implemented the above procedure based on the PySCF quantum chemistry package~\cite{sun2018pyscf,sun2020recent}. 
For the SOC treatment, we used the ``exact two-component'' Hamiltonian including only the one-electron term (including the one-electron spin-orbit coupling but not including the two-electron spin-same-orbit and spin-other-orbit coupling~\cite{li2014spin,liu2009exact}). For the systems we study below (involving Er, Dy, Ho), the lowest $|JM\rangle$ multiplet is energetically well separated and corresponds to a maximal $J$ for the ground-state $f$-shell configuration. We first converged the generalized Kohn-Sham calculations without any constraint to find the lowest energy spin-orbit coupled mean-field solution. 
We then applied spin- and orbital-rotation operators to rotate the $\vec{J}$ vector to generate a new initial guess and minimized the energy with the penalty term $\lambda (\vec{J}\cdot \vec{n} - J)$ for a given $\vec{n}$, using $\lambda=0.1 E_h$. The set of projection axes were chosen to be Haar random \REV{(uniform distribution)}, and the generalized Kohn-Sham energies were converged to $10^{-9}$ $E_h$. For a fixed target vector $\vec{n}$ and appropriate functionals (see below), we found that this procedure converges to constrained determinants within the ground spin-orbit manifold.

Given these GKS solutions, we evaluated their energies and $C_{JM,i}$ coefficients as outlined above and fitted the crystal field parameters $B_k^q$ to Eq.~\ref{eq:stevens}.
We only included $B_k^q$ up to order $k=6$ as the higher order terms are conventionally ignored. The matrix forms of the operators $O_k^q$ in the $\ket{JM}$ basis were calculated with the PyCCE program~\cite{onizhuk2021pycce}.

\begin{sloppypar}
The structures of Er-trensal (H$_3$trensal = $2,2',2''$-tris(salicylideneimino)triethylamine) (\textbf{1})~\cite{ungur2017ab}, Cs$_2$NaDyCl$_6$  (\textbf{2})~\cite{goff2010synthesis}, (C(NH$_2$)$_3$)$_5$[Er(CO$_3$)$_4$]$\cdot$11H$_2$O (\textbf{3})~\cite{aravena2016periodic},
[(Cp$^{i\mathrm{Pr}5}$)Dy(Cp*)]$^{+}$ (Cp$^{i\mathrm{Pr}5}$ = penta-iso-propylcyclopentadienyl; Cp* = pentamethylcyclopentadienyl) (\textbf{4})~\cite{ullah2019silico}, and [HoPc$_2$]$^{-}$ 
(\textbf{5})~\cite{marx2014spectroscopic} were taken from the literature. The $z$ axes of $\hat{H}_\mathrm{CF}$ were defined as the pseudo-$C_3$ symmetry axis, one of the three $C_4$ axes in the $O_h$ symmetry group, the pseudo-$C_{2v}$ symmetry axis, and the main magnetization axis of the ground Kramers doublet calculated at the CASSCF level with an $f$-shell active space (because the molecule has no symmetry), 
and the pseudo-$C_4$ symmetry axis, respectively. 
\end{sloppypar}

In the generalized Kohn-Sham calculations, we used the segmented all-electron relativistically contracted basis of valence triple-zeta quality (SARC-TZV) basis~\cite{pantazis2009all} for the lanthanides and the 6-31G basis~\cite{ditchfield1971a,hehre1972a,francl1982a, gordon1982a} for the other elements. As discussed in the results, we performed calculations using pure and hybrid density functionals, namely  PBE~\cite{perdew1996generalized}, B97-D~\cite{grimme2006semiempirical}, TPSS~\cite{tao2003climbing}, r$^2$SCAN~\cite{furness2020accurate},B3LYP~\cite{becke1988density,lee1988development,becke1993new}, PBE0\cite{perdew1996rationale}, M06\cite{zhao2008m06}. It is well known that the ground-spin-state of a metal ion can be strongly influenced by the choice of functional~\cite{reviakine_calculation_2006,ye2010accurate,reiher2001reparameterization,harvey2006accuracy,swart2016spinning}. In these systems we found that the choice of functional can have a significant impact on the effective $J$ state of the ground-state ion, as seen in Table~\ref{table:ground_J}. In particular, we found that although density functionals (other than, perhaps r$^2$SCAN) give reasonable effective $L$ \REV{(orbital angular momentum quantum number)} and $S$ \REV{(spin quantum number)} values for the central ion, most do not give ground-state $J$ values consistent with Hund's rule (which here would mean that $J = L + S$). In fact, only the Hartree-Fock functional yielded a ground-state with an overwhelming weight in the Hund's rule $|JM\rangle$ manifold (which is expected to be a well isolated manifold for the ions of interest due to the strong spin-orbit coupling).
\begin{table}   \captionsetup{singlelinecheck=false}
  \centering
  \begin{tabular}{>{\centering\arraybackslash}p{2cm}>{\centering\arraybackslash}p{1.3cm}>{\centering\arraybackslash}p{1.3cm}>{\centering\arraybackslash}p{1.3cm}>{\centering\arraybackslash}p{1.3cm}>{\centering\arraybackslash}p{1.3cm}>{\centering\arraybackslash}p{1.3cm}>{\centering\arraybackslash}p{1.3cm}}
    \toprule
           Functional & HF & M06 & PBE0 & B97-D & B3LYP & PBE & r2SCAN  \\ \midrule
   J & 7.5 & 7.3 & 7.1 & 7.1 & 7.1 & 6.7 & 5.9 \\
   L & 5.1 & 5.1 & 5.1 & 5.1 & 5.1 & 4.9 & 4.6 \\
   S & 2.4 & 2.4 & 2.4 & 2.3 & 2.4 & 2.3 & 2.4 \\
   \bottomrule 
  \end{tabular}  
  \caption{The approximate J quantum number as defined by $J(J+1)=\langle J^2\rangle$ and similarly defined L and S of the lanthanide $f$-shells in the ground state wavefunction of compound \textbf{2}, calculated using HF and various DFT functionals. }
  \label{table:ground_J}
\end{table}

Consequently, we also performed calculations where the Slater determinant was determined using generalized Hartree-Fock theory, and the energies were subsequently computed using a different functional, to avoid the above unphysical nature of the DFT ground-state, and in the spirit of minimizing density driven errors in DFT {calculations~\cite{vuckovic2019density}}. In the latter case, we refer to the calculation as $X @ Y$ where $X$ is the method used to generate the density matrix, and $Y$ is the method use to evaluate the energy. 

For each compound, 80 target orientations were sampled, except for compound \textbf{5} where 120 orientations were used. We estimate that sampling 80 orientations was sufficient to achieve a standard deviation in the relative eigenstate energies of $\hat{H}_\text{CF}$ of less than 5 cm$^{-1}$ \REV{for compound \textbf{1} (Figure~\ref{figure:precision_ErT}), though fewer samples are required for systems with higher symmetry. For example, in compound \textbf{2}, as few as 10 orientations would suffice (Figure~\ref{figure:precision_DyCl6}). Moreover, the sampling adds negligible wall time because the calculations are embarrassingly parallel.}

For the Ho compound 
 (compound \textbf{5}) we generated additional benchmark data using CASSCF with spin-orbit coupling included via state-interaction (CASSCF-SISO) and CASSCF/NEVPT2-SISO, which additionally includes dynamical correlations through NEVPT2,  using ORCA version 5.0.4~\cite{neese2012orca,ORCA5} and subsequently, the effective crystal field Hamiltonians were calculated with the SINGLE\_ANISO program~\cite{chibotaru2012ab, ungur2017ab}. In the following, we will refer to the two approaches as CASSCF and NEVPT2, respectively, and similarly refer to CASSCF/CASPT2-SISO as CASPT2. In these calculations, the second-order scalar relativistic Douglas–Kroll–Hess (DKH) Hamiltonian~\cite{DK1974,Hess1985} was used for Ho and the second-order DKH transformation was applied to the one-electron part of the SOC operator~\cite{Neese2005DKH,Sandhoefer2012DKH} with the two-electron part of the SOC operator treated by the spin-orbit mean field (SOMF) approximation.~\cite{HebetaSOMF1996}
The SARC2-DKH-QZVP basis set~\cite{Aravena2016} was used for the Ho atom, and the DKH-def2-TZVP basis set~\cite{Pantazis2008,Weigend2005} was used for all the other atoms. The spin-orbit interaction was included via quasi-degenerate perturbation theory (QDPT), a type of state interaction method~\cite{Malmqvist1989,Malmqvist2002}, 
using 35 lowest-energy quintet roots, 106 lowest-energy triplet roots, 31 lowest-energy singlet roots. 

Finally, for compound \textbf{5}, the finite temperature field-dependent magnetization was simulated with the program PHI~\cite{chilton2013phi}.

For compounds \textbf{2} and \textbf{3}, only the central clusters Er(CO$_3$)$_4$ and [DyCl$_6$]$^-$ were treated quantum mechanically. The clusters were embedded in  infinite lattices of classical charges of the remaining atoms to represent the crystal environment~\cite{li2024accurate}. The classical charges were derived from the Bader population analysis of the $\Gamma$-point PBE calculation of the lattices using the VASP program~\cite{kresse1993ab,kresse1994ab,kresse1996efficiency,kresse1996efficient} with periodic boundary conditions applied to the primitive unit cells and with the energy converged to $10^{-8}$ eV per cell. These atomic charges were then modeled as Gaussian-distributed charges on the remaining atoms using ionic radii~\cite{shannon1976revised} for Er$^{3+}$ and Dy$^{3+}$ and covalent radii~\cite{pyykko2009molecular} for the other atoms.

\paragraph{Results and discussion} 

We now discuss the constrained DFT ab initio $\hat{H}_\mathrm{CF}$ derived for  five single-ion lanthanide magnets: Er-trensal (\textbf{1}), Cs$_2$NaDyCl$_6$ (\textbf{2}), (C(NH$_2$)$_3$)$_5$[Er(CO$_3$)$_4$]·11H$_2$O (\textbf{3}), [(Cp$^{i\mathrm{Pr}5}$)Dy(Cp*)]$^+$ (\textbf{4}), and [HoPc$_2$]$^{-}$ (\textbf{5}). We compare our results to the
energies and wavefunctions obtained by diagonalizing model Hamiltonians 
in the literature, 
obtained from experiment or multi-configurational methods, and in the case of compound \textbf{5}, to additional data computed in this work using multi-configurational methods.

\begin{figure} 
    \centering
    \includegraphics[width=0.5\columnwidth]{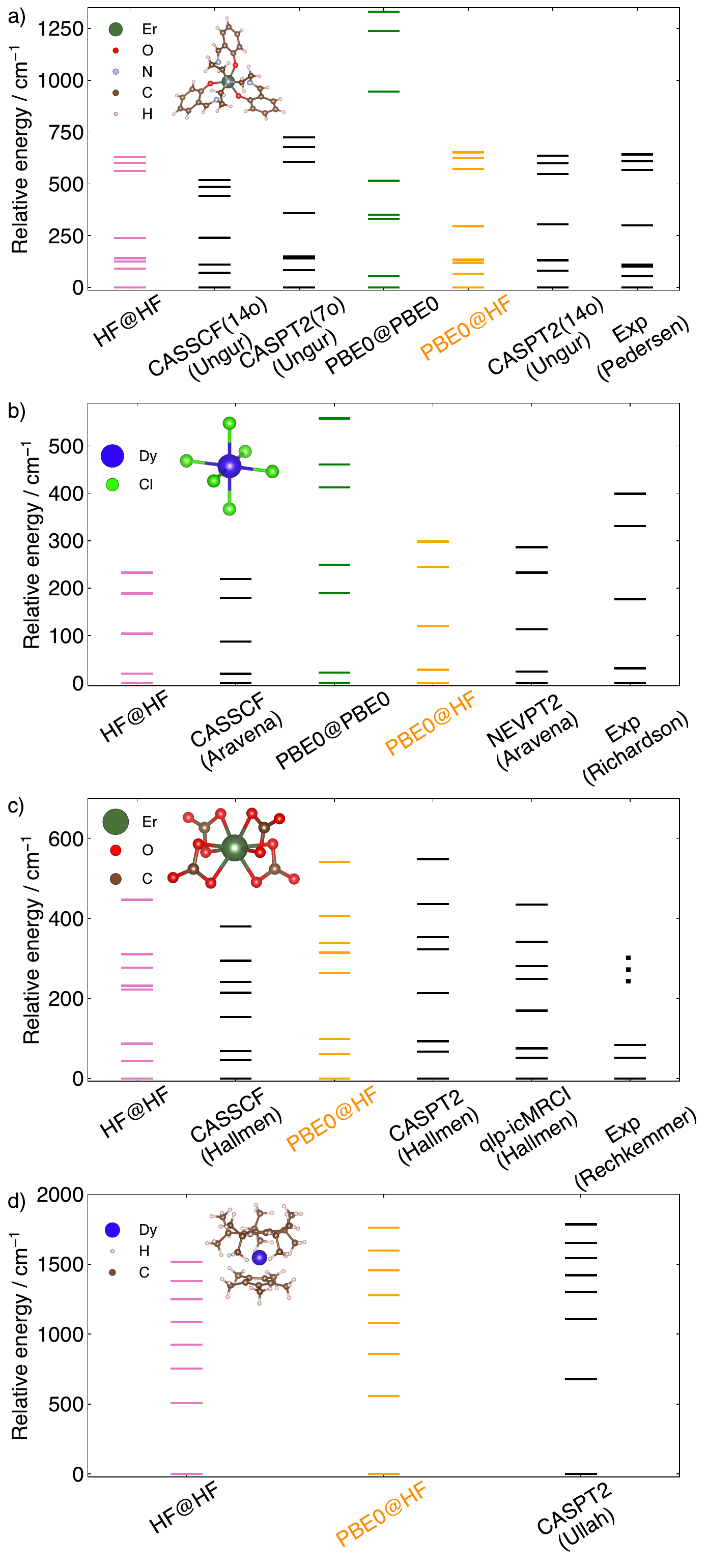}
    \caption{Relative energies of ground and low-energy excited states of \textbf{1}--\textbf{4} (\textbf{a}--\textbf{d}) from $\hat{H}_\mathrm{CF}$ using the constrained DFT methodology in this work. Results are shown for three choices of functionals: HF@HF (pink), PBE0@PBE0 (green), and PBE0@HF (orange). We compare to spectra from literature CASSCF, CASPT2 or NEVPT2, and  experiments (black)~\cite{ungur2017ab,pedersen2014modifying,flanagan2001ligand,aravena2016periodic,richardson1985energy,hallmen2018crystal,rechkemmer2015comprehensive,ullah2019silico}. }
    \label{figure:cf_cas}
\end{figure}

System \textbf{1}~\cite{kanesato1999synthesis} is a single-ion magnet (SIM) that has been extensively studied both experimentally~\cite{flanagan2001ligand,pedersen2014modifying} and theoretically~\cite{pedersen2014modifying,ungur2017ab}. It has an approximate $C_3$ symmetry and an Er $^4I_{7.5}$ ground state. Due to the time-reversal symmetry, each eigenstate is doubly degenerate, forming in total 8 Kramers doublets (KDs) in the ground-state $J=7.5$ manifold. Figure~\ref{figure:cf_cas}\textbf{a} shows the energies of the 16 states from the constrained DFT ab initio crystal field Hamiltonian using the HF, PBE0~\cite{perdew1996rationale}
and PBE0@HF mean-field functionals, and from theory and experiments in the literature. The HF-derived Hamiltonian yields a spectrum  comparable to that of CASSCF, while the PBE0@HF derived $\hat{H}_\text{CF}$ yields a spectrum closer to that from the much more expensive XMS-CASPT2 method using the  11 electrons in 14 orbitals active space. In fact, the PBE0@HF $\hat{H}_\text{CF}$  and the XMS-CASPT2 (11e, 14o active space) spectrum are in best agreement with the experimentally derived spectrum. 
We note that using DFT functionals (beyond HF) to include dynamical correlation in our procedure is required to correctly order the closely-spaced  3$^\mathrm{rd}$ and 4$^\mathrm{th}$ KDs, and that similarly out of the wavefunction methods only XMS-CASPT2 (11e, 14o active space) obtains the correct ordering of these states.
However, the density-driven error effect of DFT can be significant. For example, deriving $\hat{H}_\text{CF}$ using the PBE0 density matrix and PBE0 energy leads to worse results, due to the unphysical nature of the ground-state observed above. 
In general, we find that
neglecting dynamical correlation in the modeling (e.g., by using HF energies to derive $\hat{H}_\text{CF}$ or by obtaining the spectrum from CASSCF) tends to predict too narrow of a spectral spread
while dynamical correlation (e.g., using our PBE0@HF energies, or from literature CASPT2 results) results in a larger spread, more comparable to that of experiment~\cite{ungur2017ab,aravena2016periodic}.

\begin{figure}  
    \centering
    \includegraphics[width=0.6\columnwidth]{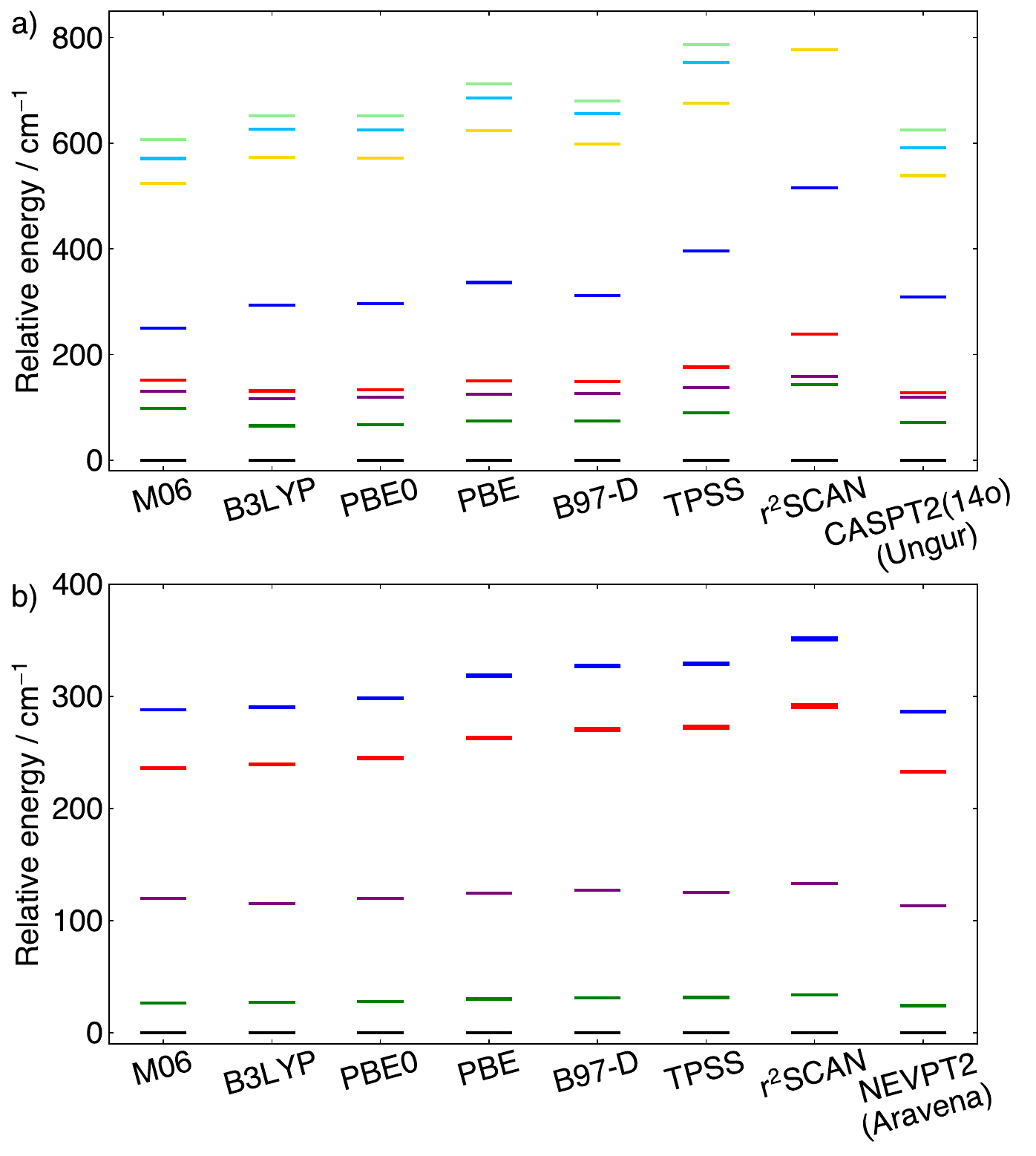}
    \caption{Relative eigenstate energies of \textbf{a}) \textbf{1} and \textbf{b}) \textbf{2} from $\hat{H}_\text{CF}$ derived from DFT@HF sampled energies with various DFT functionals~\cite{zhao2008m06,becke1988density,lee1988development,becke1993new,perdew1996rationale,grimme2006semiempirical,tao2003climbing,furness2020accurate}. We compare to the best available multireference perturbation theory calculations using CASPT2~\cite{ungur2017ab} or NEVPT2 ~\cite{aravena2016periodic} (last column). Each color represents a doublet in \textbf{a}) and an energy level with ideally 2 or 4 degeneracy in \textbf{b}).}
    \label{figure:functional_main}
\end{figure}

Similar observations apply to complexes $\textbf{2}$-\textbf{4}: the ab initio crystal field Hamiltonian derived from Hartree-Fock energies produces a spectrum comparable to the literature CASSCF derived spectra, while
 using DFT energies evaluated for the Hartree-Fock density matrices, such as when using PBE0@HF to derive $\hat{H}_\text{CF}$, yields (for most functionals) spectra comparable to the best multi-reference perturbation theory results in the literature, with similarly good agreement with experiment where available (see Figure~\ref{figure:cf_cas}, \ref{figure:functional_main}, \ref{figure:functional_supp1}, and \ref{figure:functional_supp2}).

\begin{figure} 
    \centering
    \includegraphics[width=0.6\columnwidth]{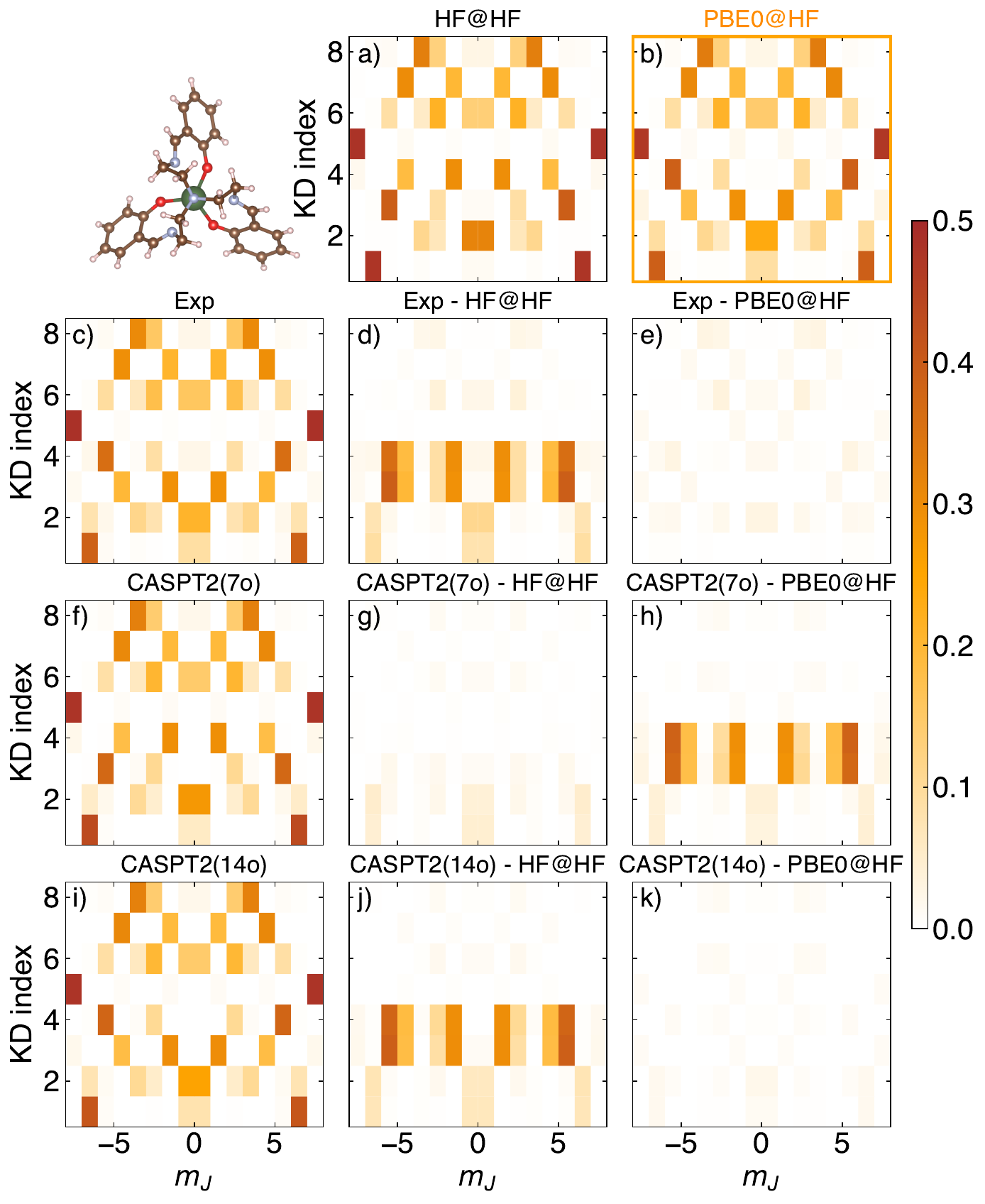}
    \caption{$\ket{JM}$ compositions,  $|c_{JM}|^2$ where $c_{JM} = \braket{JM}{\Psi}$, of each Kramers doublet (KD) of \textbf{1} from diagonalizing $\hat{H}_\mathrm{CF}$. 
    Each panel corresponds to a different $\hat{H}_\mathrm{CF}$, derived from the constrained DFT methodology using the HF@HF functional (``HF@HF'') and PBE0@HF functional (``PBE0@HF''), from
    experimental measurements~\cite{pedersen2014modifying} (``Exp''), from CASPT2 in a 7-orbital active space (``CASPT2(7o)'')~\cite{ungur2017ab}, and from CASPT2 in a 14-orbital active space (``CASPT2(14o)'')~\cite{ungur2017ab}, as well as differences between pairs. $\ket{JM}$ compositions of a KD are calculated as the average $|c_{JM}|^2$ of the two corresponding degenerate eigenstates.}
    \label{figure:wf}
\end{figure}

We observe that our ab initio $\hat{H}_\text{CF}$ also accurately predicts the $\ket{JM}$ composition of the crystal field eigenstates. We consider, for example, compound \textbf{1}. As shown in Figure~\ref{figure:wf}, the $\hat{H}_\text{CF}$ derived from the PBE0@HF energies \REV{(Figure~\ref{figure:wf}\textbf{b})} yields $\ket{JM}$ compositions for the 8 Kramers doublets that deviate from the experimentally derived compositions \REV{(Figure~\ref{figure:wf}\textbf{c})} by
only 0.6\% on average \REV{(Figure~\ref{figure:wf}\textbf{e} for the deviation)}.
To obtain a similar accuracy in a multiconfigurational calculation requires using a second-order perturbative correction and a large active space of 14o \REV{(Figure~\ref{figure:wf}\textbf{i})} as neither CASSCF(14o) nor CASPT2(7o) \REV{(Figure~\ref{figure:wf}\textbf{f})} capture the correct energy ordering.

Lastly, we investigate a larger complex, the holmium double decker compound \textbf{5}~\cite{ishikawa2003lanthanide}.  Since its discovery two decades ago as one of the first lanthanide SIMs~\cite{ishikawa2003lanthanide}, multiple experiments and calculations have sought to understand the relative energies and wavefunctions of the ground-state $J$ manifold~\cite{ishikawa2003lanthanide, Baldovi2012modeling, marx2014spectroscopic}, but have arrived at different conclusions.
The primary challenge in this compound is that the size of the molecule makes it computationally challenging to apply CASPT2 or NEVPT2 levels of theory in conjunction with (potentially necessary) large active spaces and large basis sets.

\begin{figure} 
    \centering
    \includegraphics[width=0.8\columnwidth]{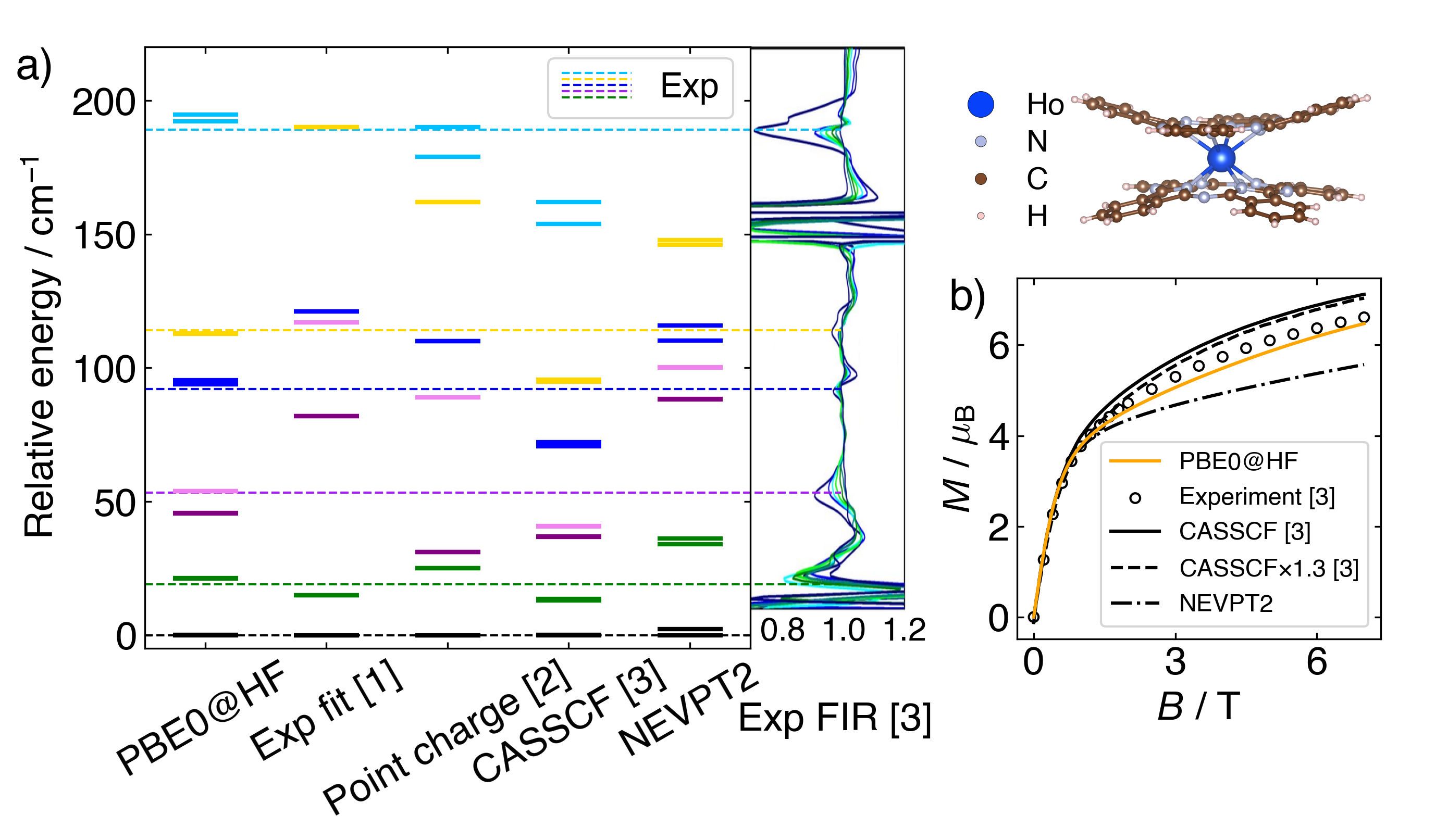}
    \caption{\textbf{a}) Relative energies of ground and low-energy excited states of \textbf{5} from $\hat{H}_\mathrm{CF}$ fitted to HF determinants and PBE0 energies (``PBE0@HF''), from $\hat{H}_\mathrm{CF}$ fitted to NMR and susceptibility experiments~\cite{ishikawa2003lanthanide} (``Exp fit [1]''), from a point charge model~\cite{Baldovi2012modeling} (``Point charge [2]''), from CASSCF in an active space of (10e, 14o)~\cite{marx2014spectroscopic} (``CASSCF [3]''), and from NEVPT2 in an active space of (10e, 7o), in comparison with the experimental FIR spectrum~\cite{marx2014spectroscopic} (``Exp FIR [3]''). The energy levels are grouped into quasi-doublets, with each pair--along with the corresponding experimental peaks (dashed lines)--colored consistently according to their energy order. The light and dark purple quasi-doublets are both assigned to the broad FIR peak near 51 cm$^{-1}$.  The peak near 150 cm$^{-1}$ in the experimental FIR spectrum is likely phononic instead of magnetic and thus ignored. Experimental FIR spectrum is reproduced with permission from the reference by Marx, et al.~\cite{marx2014spectroscopic}.  \textbf{b}) Calculated magnetization (``PBE0@HF'') as a function of the applied field at 1.8~K on a powder sample of \textbf{5}, in comparison with the literature~\cite{marx2014spectroscopic}: the experimental measurement (``Experiment [3]''), the prediction from CASSCF as in \textbf{a}) (``CASSCF [3]''), the prediction after the CASSCF relative energies are scaled by 1.3 (``CASSCF$\times$1.3 [3]'')\REV{, and that from NEVPT2 (``NEVPT2'')}. }
    \label{figure:DDeck_en}
\end{figure}

However, the computational efficiency of our method allows it to be readily applied.
Figure~\ref{figure:DDeck_en} compares the excitation energies predicted using $\hat{H}_\mathrm{CF}$ derived using our method and PBE0@HF energies, compared to predictions from other techniques. 
As shown in Figure~\ref{figure:DDeck_en}, the excitation energies predicted by our method match the peaks in the spectrum very well. (Note that the experimental peak near 150 cm$^{-1}$ is a phonon peak and should be ignored). In contrast,
neither the ligand field model directly fitted to NMR and susceptibility experiments (``Exp fit'')~\cite{ishikawa2003lanthanide} nor the parameterized effective point charge model fit to experimental $\chi T$ curves (``Point charge'')~\cite{Baldovi2012modeling} correctly captures the second set of excited quasi-doublets (purple). The energies predicted by CASSCF (7o) must be rescaled by an empirical factor 1.3 to match the lowest three observed excited quasi-doublets in the spectrum~\cite{marx2014spectroscopic} while our method requires no empirical adjustment. We also performed NEVPT2 calculations using a 7-orbital active space, which, at the  CASSCF level, accurately reproduced the CASSCF (7o) results in the literature~\cite{marx2014spectroscopic},  and found that the excitations deviate from the experimental spectrum even more than the CASSCF results, perhaps due to the small active space, as observed previously in compound \textbf{1}. 
To further compare against experimental observables, we computed the field-dependent magnetization (Figure~\ref{figure:DDeck_en}\textbf{b}) and temperature-dependent magnetic susceptibility (Figure~\ref{figure:DDeck_chiT}) using our ab initio $\hat{H}_\text{CF}$,  where we find good agreement with the experimentally measured curves, improving on the CASSCF derived curves.
 
\paragraph{Conclusion}

In conclusion, we have developed a method to determine the 
effective crystal field Hamiltonian of single-ion magnets with quantitative accuracy at mean-field cost. The accuracy of the crystal field Hamiltonian parameters that we obtain is comparable to that of the best multi-reference perturbation theory calculations in the literature, as we have demonstrated on four benchmark literature systems. 
In addition, we showcased an application to the large holmium double decker complex whose low-energy spectrum has previously been challenging to assign theoretically. Using our mean-field derived Hamiltonian, we were able to reproduce all low-energy states measured in the experimental spectra as well as the field-dependent magnetic susceptibility curve.
The low-computational cost and high-accuracy of our approach will allow for applications to larger and more complex molecules. This method thus lays the foundation for future quantitative simulations of the low-energy physics of single-ion magnets and optimizing their design.

\section{Acknowledgments}
L.P. acknowledges the helpful discussion with Huanchen Zhai and Samuel Trickey.
This work is supported by the Center for Molecular Magnetic Quantum Materials (M2QM), an Energy Frontier Research Center funded by the US Department of Energy, Office of Science, Basic Energy Sciences under Award DE-SC0019330. The computational studies employed resources of the National Energy Research Scientific Computing Center (NERSC), a U.S. Department of Energy Office of Science User Facility located at Lawrence Berkeley National Laboratory, operated under Contract no. DE-AC02-05CH11231. 

\section{Competing interest}
G.K.C. is a part owner of QSimulate. The remaining authors declare no competing interests.

\section{Data availability}
\REV{The implementation of the theoretical method described in this work is openly available at \url{https://github.com/penglinq/cDFT-CF}. Example input and output files for reproducing the data of the systems studied in this manuscript can also be found in the same repository.}

\bibliography{references}

\clearpage

\begin{center}
\textbf{\Large Supporting information for: \\ Accurate crystal field Hamiltonians of single-ion magnets at mean-field cost}
\end{center}

\setcounter{equation}{0}
\setcounter{figure}{0}
\setcounter{table}{0}
\setcounter{page}{1}
\setcounter{section}{0}
\setcounter{secnumdepth}{3}
\renewcommand{\theequation}{S\arabic{equation}}
\renewcommand{\thefigure}{S\arabic{figure}}
\renewcommand{\thetable}{S\arabic{table}}
\renewcommand{\thesection}{S\arabic{section}}

\begin{figure} [!htb]
    \centering
    \includegraphics[width=0.75\columnwidth]{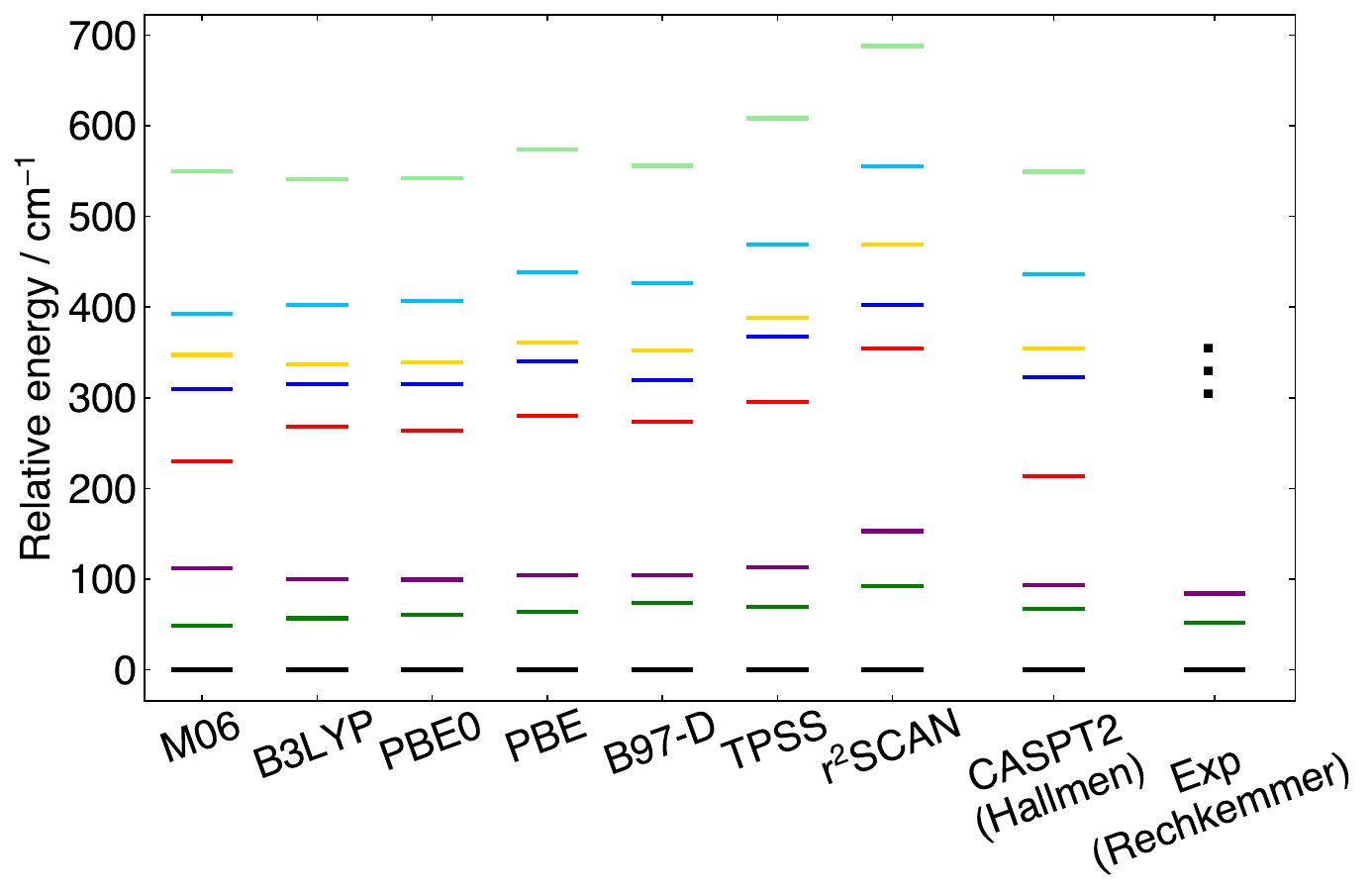}
    \caption{Relative energies of eigenstates of \textbf{3} when different DFT functionals are used to evaluate energy expectation values of cHF determinants in comparison to values from CASPT2~\cite{hallmen2018crystal} and experiments~\cite{rechkemmer2015comprehensive} (last two columns). Each color represents a degenerate doublet.}
    \label{figure:functional_supp1}
\end{figure}

\begin{figure} [!htb]
    \centering
    \includegraphics[width=0.75\columnwidth]{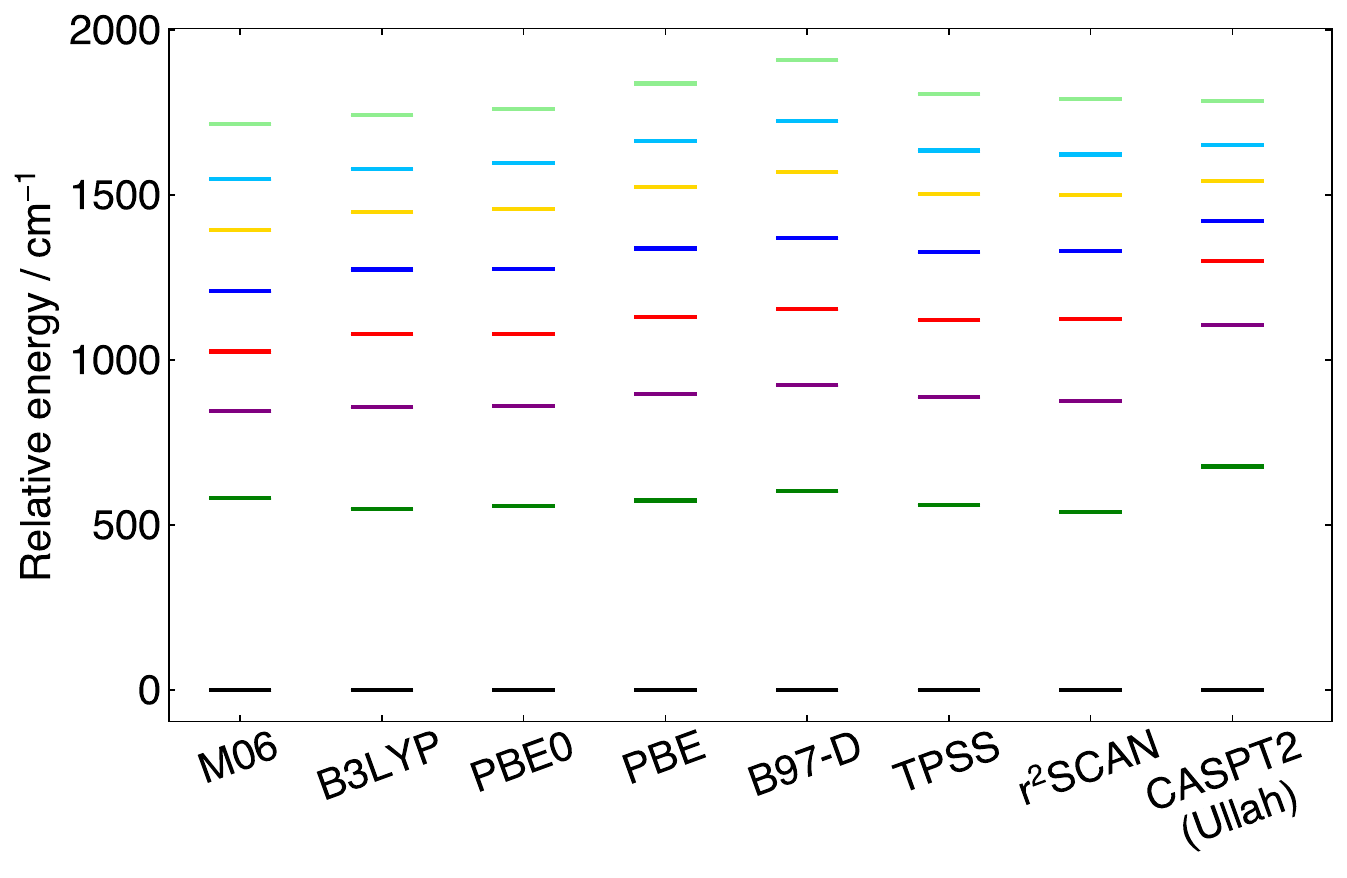}
    \caption{Relative energies of eigenstates of \textbf{4} when different DFT functionals are used to evaluate energy expectation values of cHF determinants in comparison to values calculated by CASPT2~\cite{ullah2019silico} (last column). Each color represents a degenerate doublet.}
    \label{figure:functional_supp2}
\end{figure}

\begin{figure} [!htb]
    \centering
    \includegraphics[width=0.75\columnwidth]{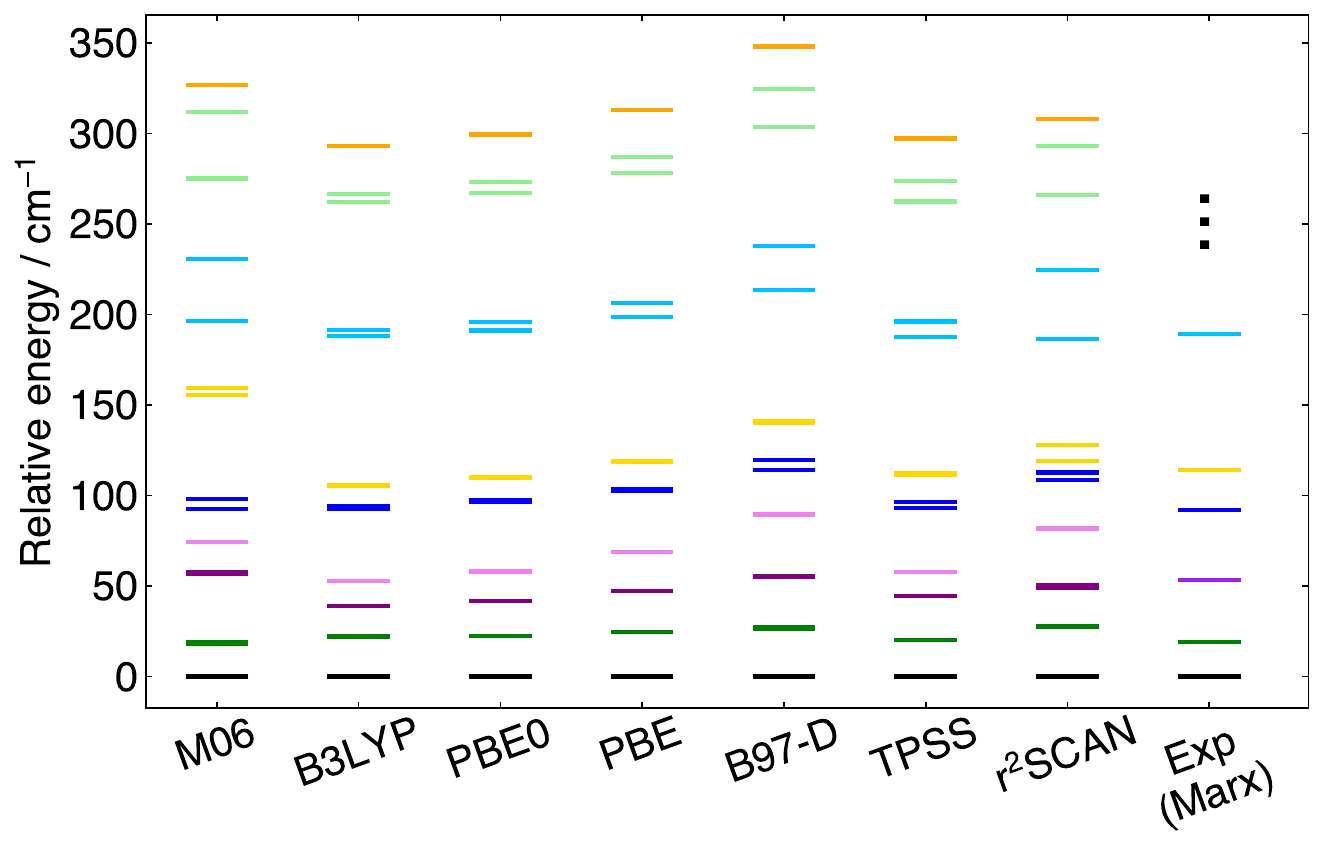}
    \caption{Relative energies of eigenstates of \textbf{5} when different DFT functionals are used to evaluate energy expectation values of cHF determinants in comparison to values extracted from the experiment~\cite{marx2014spectroscopic} (last column). Each color except black represents a quasi-doublet.}
    \label{figure:functional_supp3}
\end{figure}

\begin{figure} 
    \centering
    \includegraphics[width=0.75\columnwidth]{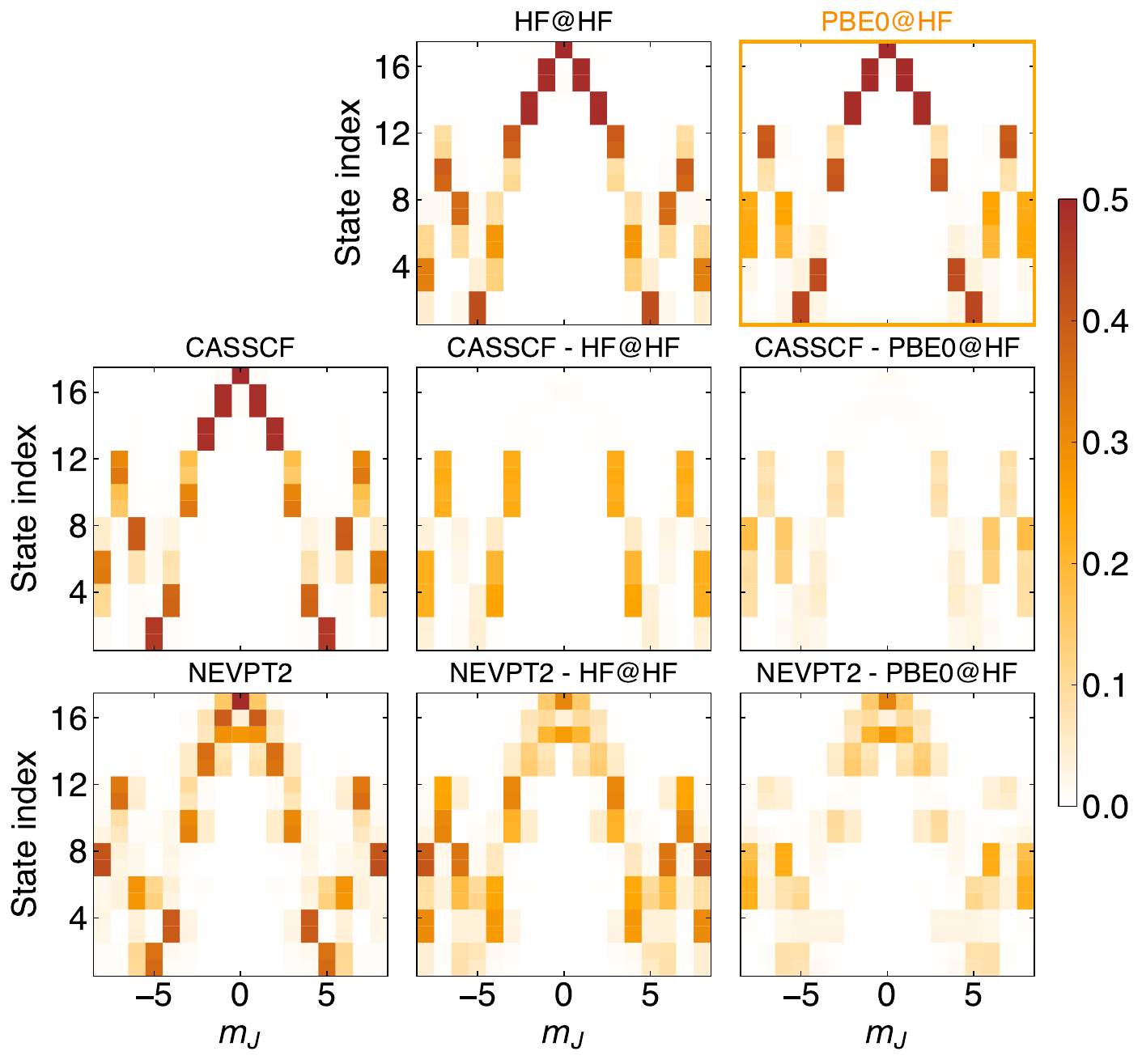}
    \caption{\REV{$\ket{JM}$ compositions,  $|c_{JM}|^2$ where $c_{JM} = \braket{JM}{\Psi}$, of each eigenstate of \textbf{5} from diagonalizing $\hat{H}_\mathrm{CF}$. Each panel corresponds to a different $\hat{H}_\mathrm{CF}$, derived from the constrained DFT methodology using HF@HF functional (``HF@HF'') and PBE0@HF functional (``PBE0@HF''), from CASSCF in an active space of (10e, 14o)(``CASSCF''), and by NEVPT2 in an active space of (10e, 7o)  (``NEVPT2''),  as well as differences between pairs. }} 
    \label{figure:wf_supp1}
\end{figure}

\begin{figure} 
    \centering
    \includegraphics[width=0.75\columnwidth]{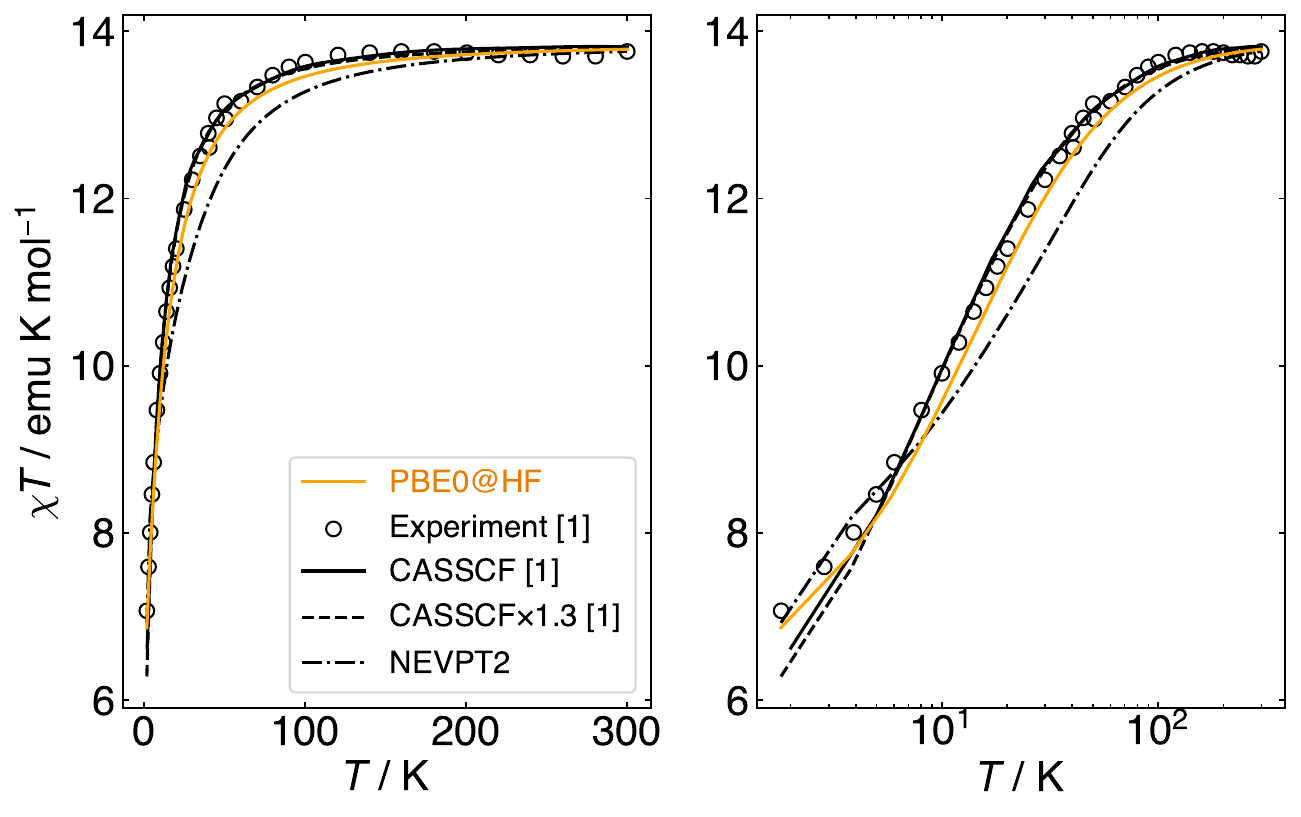}
    \caption{Calculated magnetic susceptibility (``PBE0@HF'') as a function of temperature on a powder sample of \textbf{5}, in comparison with the literature~\cite{marx2014spectroscopic}: the experimental measurement (``Experiment [1]''), the prediction from CASSCF in an active space of (10e, 14o) (``CASSCF [1]''), the prediction after the CASSCF relative energies are scaled by 1.3 (``CASSCF$\times$1.3 [1]'')\REV{, and that from NEVPT2 in an active space of (10e, 7o) (``NEVPT2'')}. Note: the main mismatch between the experiment and all theoretical calculations at high temperatures is likely due to experimental uncertainties because $\chi T$ is expected to change monotonically in the high temperature regime.   } 
    \label{figure:DDeck_chiT}
\end{figure}

\begin{figure} 
    \centering
    \includegraphics[width=0.75\columnwidth]{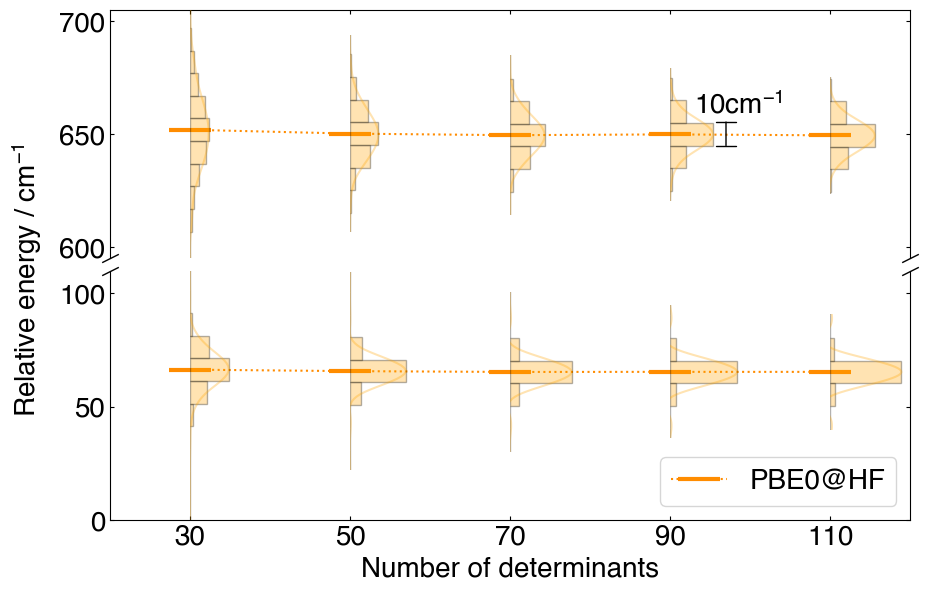}
    \caption{\REV{Average relative energies of the first and highest excited states computed from 500 selections, each of 30, 50, 70, 90, and 110 determinants (orange marker) and the corresponding relative energy distributions (orange histogram) of \textbf{1}. Because \textbf{1} has $C_3$ symmetry, 9 $B_k^q$ parameters that preserve the symmetry are fitted. About 80 determinants are enough to achieve 5 cm$^{-1}$ precision (measured by average standard deviation).}  }
    \label{figure:precision_ErT}
\end{figure}

\begin{figure} 
    \centering
    \includegraphics[width=0.75\columnwidth]{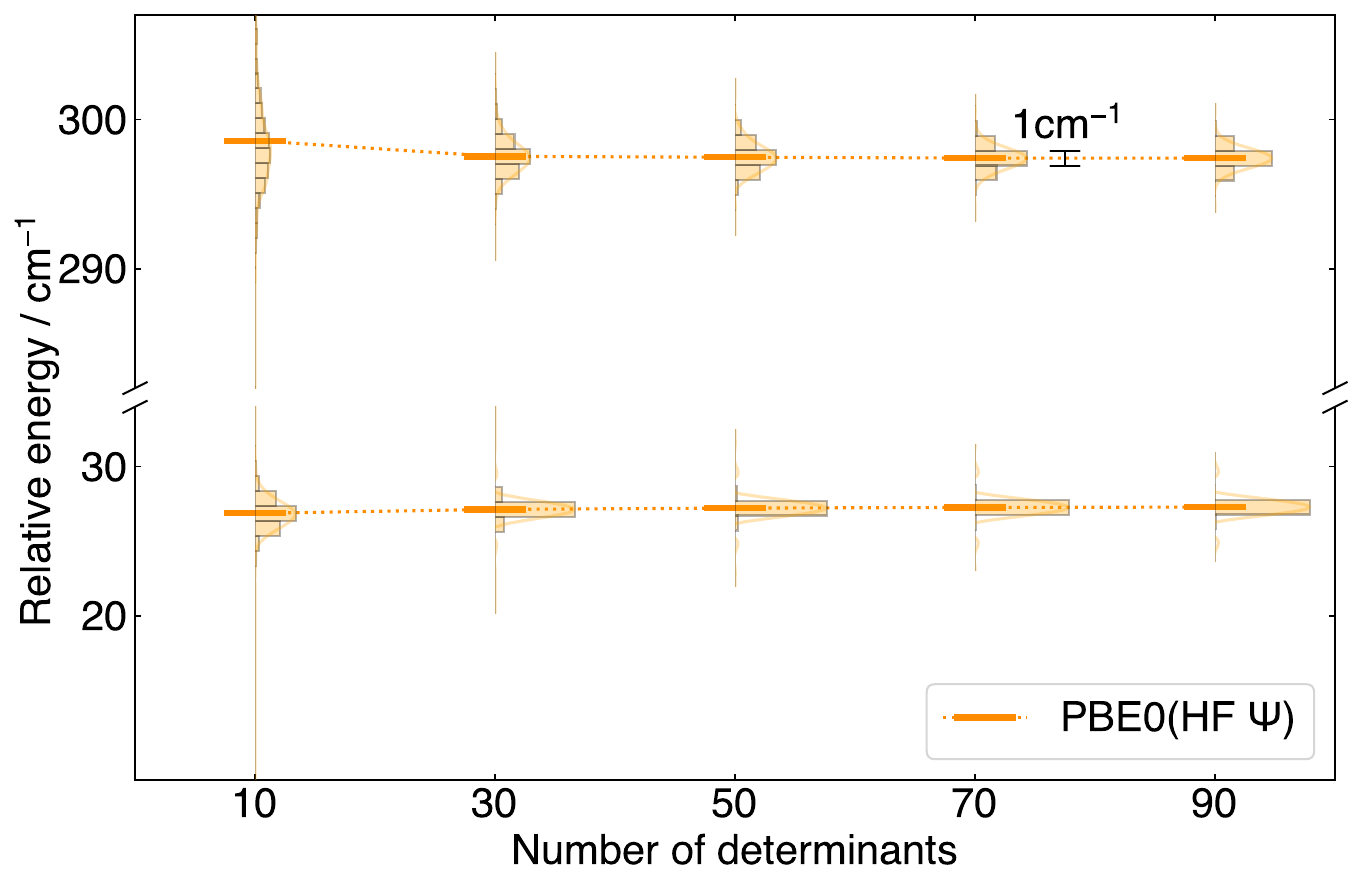}
    \caption{\REV{Average relative energies of the first and highest excited states computed from 500 selections, each of 10, 30, 50, 70, and 90 determinants (orange marker) and the corresponding relative energy distributions (orange histogram) of \textbf{2}. Because \textbf{2} has $O_h$ symmetry, 4 $B_k^q$ parameters that preserve the symmetry are fitted. 10 determinants are enough to achieve 5 cm$^{-1}$ precision (measured by average standard deviation).}  }
    \label{figure:precision_DyCl6}
\end{figure}
\clearpage

\begingroup
\renewcommand\arraystretch{0.65}
\begin{longtable}{crrrr} 
    \toprule
    \multirow{2}{*}{$k$} & \multirow{2}{*}{$q$} & \multicolumn{3}{c}{$B_k^q$ ($\mathrm{cm}^{-1}$)}  \\
    & & HF@HF & PBE0@HF & \REV{CASPT2(14o) (Ungur)} \\
    \midrule
    2 & 2 & 0 & 0 & \REV{-2.2246278E-03} \\
    2 & 1 & 0 & 0 & \REV{5.6794895E-04} \\
    2 & 0 & -1.0846603E+00 & -1.1673728E+00 & \REV{-1.1122607E+00} \\
    2 & -1 & 0 & 0 & \REV{-1.2014236E-03} \\
    2 & -2 & 0 & 0 & \REV{-2.3038309E-03} \\
    4 & 4 & 0 & 0 & \REV{-2.2169983E-04} \\
    4 & 3 & 1.6502459E-01 & 2.0880005E-01 & \REV{1.9528516E-01} \\
    4 & 2 & 0 & 0 & \REV{-1.9816623E-05} \\
    4 & 1 & 0 & 0 & \REV{6.0357855E-05} \\
    4 & 0 & -1.0765159E-03 & -1.2073254E-03 & \REV{-5.5875272E-04}\\
    4 & -1 & 0 & 0 & \REV{4.6727343E-05}\\
    4 & -2 & 0 & 0 & \REV{-2.7349370E-05}\\
    4 & -3 & -9.1485833E-02 & -1.1321827E-01 & \REV{-1.0677587E-01} \\
    4 & -4 & 0 & 0 & \REV{-5.4909958E-05}\\
    6 & 6 & 4.6114261E-04 & 6.6374329E-04 & \REV{6.7345254E-04}\\
    6 & 5 & 0 & 0 & \REV{-2.9561071E-06} \\
    6 & 4 & 0 & 0 & \REV{1.2036835E-06} \\
    6 & 3 & -6.7258782E-04 & -1.2066800E-03 & \REV{-1.0575970E-03} \\
    6 & 2 & 0 & 0 & \REV{3.4740060E-08} \\
    6 & 1 & 0 & 0 & \REV{-1.9056907E-06} \\
    6 & 0 & 9.3301951E-05 & 1.3333043E-04 & \REV{1.3367383E-04} \\
    6 & -1 & 0 & 0 & \REV{-1.4077862E-06} \\
    6 & -2 & 0 & 0 & \REV{5.6767376E-07} \\
    6 & -3 & 4.4996124E-04 & 7.6336324E-04 & \REV{6.3971423E-04} \\
    6 & -4 & 0 & 0 & \REV{4.2284782E-07} \\
    6 & -5 & 0 & 0 & \REV{-1.2967029E-05} \\
    6 & -6 & -6.6657762E-04 & -1.0271989E-03 & \REV{-9.6671484E-04}\\
    \bottomrule
  \caption{\REV{Effective} crystal field parameters $B_k^q$ in cm$^{-1}$ of \textbf{1} calculated by HF@HF and PBE0@HF \REV{compared to $B_k^q$ calculated by CASPT2 in the active space of 11 electrons in 14 orbitals from the 
  literature~\cite{ungur2017ab} in the coordinate system used by HF@HF and PBE@HF.}}
\end{longtable}
\endgroup

\clearpage
\begingroup
\renewcommand\arraystretch{0.65}
\begin{longtable}{crrr} 
    \toprule
    \multirow{2}{*}{$k$} & \multirow{2}{*}{$q$} & \multicolumn{2}{c}{$B_k^q$ ($\mathrm{cm}^{-1}$)}  \\
    & & HF@HF & PBE0@HF \\
    \midrule
    2 & 2 & 0 & 0\\
    2 & 1 & 0 & 0\\
    2 & 0 & 0 & 0\\
    2 & -1 & 0 & 0\\
    2 & -2 & 0 & 0\\
    4 & 4 & -3.3858198E-02 & -4.2870167E-02 \\
    4 & 3 & 0 & 0\\
    4 & 2 & 0 & 0\\
    4 & 1 & 0 & 0\\
    4 & 0 & -6.7731441E-03 & -8.5464965E-03 \\
    4 & -1 & 0 & 0\\
    4 & -2 & 0 & 0\\
    4 & -3 & 0 & 0\\
    4 & -4 & 0 & 0\\
    6 & 6 & 0 & 0\\
    6 & 5 & 0 & 0\\
    6 & 4 & -1.4594543E-04 & -2.5131119E-04 \\
    6 & 3 & 0 & 0\\
    6 & 2 & 0 & 0\\
    6 & 1 & 0 & 0\\
    6 & 0 & 6.8419138E-06 & 1.1472420E-05 \\
    6 & -1 & 0 & 0\\
    6 & -2 & 0 & 0\\
    6 & -3 & 0 & 0\\
    6 & -4 & 0 & 0\\
    6 & -5 & 0 & 0\\
    6 & -6 & 0 & 0\\
    \bottomrule
  \caption{\REV{Effective} crystal field parameters $B_k^q$ in cm$^{-1}$ of \textbf{2}  calculated by HF@HF and PBE0@HF.}
\end{longtable}
\endgroup

\clearpage
\begingroup
\renewcommand\arraystretch{0.65}
\begin{longtable}{crrr} 
    \toprule
    \multirow{2}{*}{$k$} & \multirow{2}{*}{$q$} & \multicolumn{2}{c}{$B_k^q$ ($\mathrm{cm}^{-1}$)}  \\
    & & HF@HF & PBE0@HF \\
    \midrule
    2 & 2 & -5.2232266E-03 & 2.8757778E-02 \\
    2 & 1 & 1.9634481E+00 & 1.8319253E+00 \\
    2 & 0 & 3.5486891E-01 & 4.4014090E-01 \\
    2 & -1 & -2.6818954E+00 & -2.5409868E+00 \\
    2 & -2 & -6.3689908E-01 & -5.9399040E-01 \\
    4 & 4 & 1.9363745E-02 & 2.0529565E-02 \\
    4 & 3 & -3.9438401E-02 & -5.2165344E-02 \\
    4 & 2 & -6.0804067E-04 & 5.7747295E-04 \\
    4 & 1 & 7.1474728E-03 & 9.6023711E-03 \\
    4 & 0 & -1.7761448E-03 & -2.0990379E-03 \\
    4 & -1 & -1.7829884E-03 & 1.8118787E-04 \\
    4 & -2 & 9.2615835E-03 & 1.1076672E-02 \\
    4 & -3 & -4.7504649E-02 & -5.9706260E-02 \\
    4 & -4 & -7.9815001E-04 & -5.7182707E-04 \\
    6 & 6 & -1.0651922E-04 & -1.5208921E-04 \\
    6 & 5 & 7.6064554E-04 & 9.0068884E-04 \\
    6 & 4 & 5.6596441E-04 & 7.6799385E-04 \\
    6 & 3 & 1.3903896E-04 & 4.0271251E-05 \\
    6 & 2 & -8.5949376E-05 & -1.1156412E-04 \\
    6 & 1 & 4.1977088E-04 & 5.5461463E-04 \\
    6 & 0 & -1.7241376E-06 & 1.2759930E-05 \\
    6 & -1 & -4.7321485E-04 & -6.1627613E-04 \\
    6 & -2 & -2.2378443E-04 & -2.7467333E-04 \\
    6 & -3 & 8.1344667E-05 & -3.3625113E-05 \\
    6 & -4 & -4.6287421E-05 & -3.2742706E-05 \\
    6 & -5 & -1.4948500E-03 & -2.0773848E-03 \\
    6 & -6 & -1.0373431E-04 & -1.0423935E-04 \\
    \bottomrule
  \caption{\REV{Effective} crystal field parameters $B_k^q$ in cm$^{-1}$ of \textbf{3} calculated by HF@HF and PBE0@HF.}
\end{longtable}
\endgroup

\clearpage
\begingroup
\renewcommand\arraystretch{0.65}
\begin{longtable}{crrr} 
    \toprule
    \multirow{2}{*}{$k$} & \multirow{2}{*}{$q$} & \multicolumn{2}{c}{$B_k^q$ ($\mathrm{cm}^{-1}$)}  \\
    & & HF@HF & PBE0@HF \\
    \midrule
    2 & 2 & -1.2287313E+01 & -1.4418493E+01 \\
    2 & 1 & -6.9757838E+00 & -8.6764100E+00 \\
    2 & 0 & 3.3649421E+00 & 3.8523165E+00 \\
    2 & -1 & 7.2530801E-01 & 9.6458528E-01 \\
    2 & -2 & -1.9844523E+00 & -2.3286842E+00 \\
    4 & 4 & -1.3786590E-03 & -1.0214740E-03 \\
    4 & 3 & -9.6165595E-03 & -1.0625839E-02 \\
    4 & 2 & 5.4788920E-03 & 6.6216666E-03 \\
    4 & 1 & 3.0796631E-03 & 3.5727372E-03 \\
    4 & 0 & -6.9378066E-04 & -7.4132671E-04 \\
    4 & -1 & 2.4932687E-03 & 2.6833707E-03 \\
    4 & -2 & -8.8646221E-04 & -1.0011731E-03 \\
    4 & -3 & -1.4583190E-03 & -7.3618383E-04 \\
    4 & -4 & -1.1634367E-03 & -9.9435934E-05 \\
    6 & 6 & -4.3117030E-04 & -4.2684506E-04 \\
    6 & 5 & -5.9581761E-04 & -5.1541134E-04 \\
    6 & 4 & 1.2975263E-04 & 9.0132116E-05 \\
    6 & 3 & 2.5312663E-04 & 2.4883638E-04 \\
    6 & 2 & -7.2431744E-05 & -5.9537346E-05 \\
    6 & 1 & -5.6990723E-05 & -4.3139328E-05 \\
    6 & 0 & 6.3312379E-06 & 5.7860428E-06 \\
    6 & -1 & -7.9778886E-06 & -1.9913231E-05 \\
    6 & -2 & -7.4718173E-06 & 1.2787499E-05 \\
    6 & -3 & 3.1009108E-05 & 3.6363419E-05 \\
    6 & -4 & 1.2464284E-04 & 1.4948835E-04 \\
    6 & -5 & -1.2528591E-04 & -6.7012911E-05 \\
    6 & -6 & -2.3136738E-04 & -2.1580377E-04 \\
    \bottomrule
  \caption{\REV{Effective} crystal field parameters $B_k^q$ in cm$^{-1}$ of \textbf{4} calculated by HF@HF and PBE0@HF.}
\end{longtable}
\endgroup

\clearpage
\begingroup
\renewcommand\arraystretch{0.65}
\begin{longtable}{crrr} 
    \toprule
    \multirow{2}{*}{$k$} & \multirow{2}{*}{$q$} & \multicolumn{2}{c}{$B_k^q$ ($\mathrm{cm}^{-1}$)}  \\
    & & HF@HF & PBE0@HF \\
    \midrule
    2 & 2 & 1.3247235E-01 & 1.0874954E-01 \\
    2 & 1 & 6.9719131E-02 & 9.9240276E-02 \\
    2 & 0 & -9.4297012E-01 & -8.6617784E-01 \\
    2 & -1 & -7.1759360E-02 & -1.7988066E-01 \\
    2 & -2 & 5.5339979E-02 & 4.4524211E-02 \\
    4 & 4 & 1.0214938E-03 & 6.7995693E-04 \\
    4 & 3 & 1.3096690E-03 & 9.3650412E-04 \\
    4 & 2 & -1.9157154E-04 & 1.5404399E-04 \\
    4 & 1 & 1.2923476E-03 & 8.9219231E-04 \\
    4 & 0 & 3.5847312E-03 & 4.3206015E-03 \\
    4 & -1 & -3.1961852E-04 & -1.0904990E-03 \\
    4 & -2 & -6.9579179E-05 & -1.8108109E-04 \\
    4 & -3 & -2.6759864E-05 & -2.9909259E-04 \\
    4 & -4 & -7.8698659E-04 & -1.1475216E-03 \\
    6 & 6 & 4.9631342E-06 & -4.4809715E-06 \\
    6 & 5 & -5.1175311E-06 & 9.2908806E-05 \\
    6 & 4 & 3.0346879E-05 & 1.6163801E-05 \\
    6 & 3 & 2.2262395E-05 & 3.5997317E-07 \\
    6 & 2 & -1.0055515E-05 & -2.7661814E-05 \\
    6 & 1 & -2.8179552E-05 & -2.0483221E-05 \\
    6 & 0 & -3.0836658E-05 & -3.6827041E-05 \\
    6 & -1 & 9.3245157E-06 & 2.7623707E-05 \\
    6 & -2 & -3.3177020E-06 & 2.1730746E-06 \\
    6 & -3 & -1.4568335E-05 & -3.1934002E-05 \\
    6 & -4 & -2.9868374E-05 & -2.0383448E-05 \\
    6 & -5 & -4.9303094E-05 & -7.8266030E-05 \\
    6 & -6 & 5.9953167E-06 & 2.0568980E-06 \\
    \bottomrule
  \caption{\REV{Effective} crystal field parameters $B_k^q$ in cm$^{-1}$ of \textbf{5} calculated by HF@HF and PBE0@HF.}
\end{longtable}
\endgroup

\clearpage
\begingroup
\renewcommand\arraystretch{0.65}
\begin{longtable}{crr} 
    \toprule
    \multirow{2}{*}{State index} & \multicolumn{2}{c}{$\Delta E$ ($\mathrm{cm}^{-1}$)}  \\
    & HF@HF & PBE0@HF \\
    \midrule
1 & 0.00 & 0.00 \\
2 & 0.00 & 0.00 \\
3 & 89.85 & 67.12 \\
4 & 89.85 & 67.12 \\
5 & 124.25 & 119.12 \\
6 & 124.25 & 119.12 \\
7 & 141.19 & 133.66 \\
8 & 141.19 & 133.66 \\
9 & 238.33 & 295.59 \\
10 & 238.33 & 295.59 \\
11 & 561.86 & 572.08 \\
12 & 561.86 & 572.08 \\
13 & 602.17 & 625.49 \\
14 & 602.17 & 625.49 \\
15 & 628.83 & 651.87 \\
16 & 628.83 & 651.87 \\
    \bottomrule
  \caption{Energy levels in cm$^{-1}$ in the ground spin-orbit manifold of \textbf{1} calculated by HF@HF and PBE0@HF.}
\end{longtable}
\endgroup

\begingroup
\renewcommand\arraystretch{0.65}
\begin{longtable}{crr} 
    \toprule
    \multirow{2}{*}{State index} & \multicolumn{2}{c}{$\Delta E$ ($\mathrm{cm}^{-1}$)}  \\
    & HF@HF & PBE0@HF \\
    \midrule
1 & 0.00 & 0.00 \\
2 & 0.00 & 0.00 \\
3 & 19.92 & 27.61 \\
4 & 19.92 & 27.61 \\
5 & 19.97 & 27.68 \\
6 & 19.97 & 27.68 \\
7 & 103.88 & 119.43 \\
8 & 103.88 & 119.43 \\
9 & 188.56 & 244.52 \\
10 & 188.56 & 244.52 \\
11 & 188.75 & 245.46 \\
12 & 188.75 & 245.46 \\
13 & 232.65 & 297.65 \\
14 & 232.65 & 297.65 \\
15 & 232.77 & 298.19 \\
16 & 232.77 & 298.19 \\
    \bottomrule
  \caption{Energy levels in cm$^{-1}$ in the ground spin-orbit manifold of \textbf{2} calculated by HF@HF and PBE0@HF.}
\end{longtable}
\endgroup

\begingroup
\renewcommand\arraystretch{0.65}
\begin{longtable}{crr} 
    \toprule
    \multirow{2}{*}{State index} & \multicolumn{2}{c}{$\Delta E$ ($\mathrm{cm}^{-1}$)}  \\
    & HF@HF & PBE0@HF \\
    \midrule
1 & 0.00 & 0.00 \\
2 & 0.00 & 0.00 \\
3 & 44.74 & 60.31 \\
4 & 44.74 & 60.31 \\
5 & 86.97 & 99.42 \\
6 & 86.97 & 99.42 \\
7 & 222.52 & 263.69 \\
8 & 222.52 & 263.69 \\
9 & 232.04 & 314.74 \\
10 & 232.04 & 314.74 \\
11 & 276.88 & 338.75 \\
12 & 276.88 & 338.75 \\
13 & 310.91 & 407.08 \\
14 & 310.91 & 407.08 \\
15 & 447.16 & 542.23 \\
16 & 447.16 & 542.23 \\
    \bottomrule
  \caption{Energy levels in cm$^{-1}$ in the ground spin-orbit manifold of \textbf{3} calculated by HF@HF and PBE0@HF.}
\end{longtable}
\endgroup

\begingroup
\renewcommand\arraystretch{0.65}
\begin{longtable}{crr} 
    \toprule
    \multirow{2}{*}{State index} & \multicolumn{2}{c}{$\Delta E$ ($\mathrm{cm}^{-1}$)}  \\
    & HF@HF & PBE0@HF \\
    \midrule
1 & 0.00 & 0.00 \\
2 & 0.00 & 0.00 \\
3 & 506.93 & 557.00 \\
4 & 506.93 & 557.00 \\
5 & 753.92 & 860.13 \\
6 & 753.92 & 860.13 \\
7 & 923.89 & 1077.78 \\
8 & 923.89 & 1077.78 \\
9 & 1088.86 & 1276.57 \\
10 & 1088.86 & 1276.57 \\
11 & 1250.01 & 1457.35 \\
12 & 1250.01 & 1457.35 \\
13 & 1379.76 & 1596.50 \\
14 & 1379.76 & 1596.50 \\
15 & 1518.41 & 1760.33 \\
16 & 1518.41 & 1760.33 \\
    \bottomrule
  \caption{Energy levels in cm$^{-1}$ in the ground spin-orbit manifold of \textbf{4} calculated by HF@HF and PBE0@HF.}
\end{longtable}
\endgroup

\begingroup
\renewcommand\arraystretch{0.65}
\begin{longtable}{crr} 
    \toprule
    \multirow{2}{*}{State index} & \multicolumn{2}{c}{$\Delta E$ ($\mathrm{cm}^{-1}$)}  \\
    & HF@HF & PBE0@HF \\
    \midrule
1 & 0.00 & 0.00 \\
2 & 0.14 & 0.16 \\
3 & 12.47 & 21.18 \\
4 & 12.52 & 21.42 \\
5 & 34.90 & 45.57 \\
6 & 35.14 & 45.68 \\
7 & 45.40 & 53.89 \\
8 & 45.56 & 53.93 \\
9 & 91.88 & 93.74 \\
10 & 92.62 & 95.28 \\
11 & 124.03 & 112.74 \\
12 & 126.83 & 112.92 \\
13 & 223.47 & 192.16 \\
14 & 233.32 & 194.69 \\
15 & 310.46 & 268.40 \\
16 & 321.22 & 273.24 \\
17 & 349.27 & 300.40 \\
    \bottomrule
  \caption{Energy levels in cm$^{-1}$ in the ground spin-orbit manifold of \textbf{5} calculated by HF@HF and PBE0@HF.}
\end{longtable}
\endgroup

\end{document}